\documentclass[epj]{svjour}
\usepackage{graphics}
\begin{document}
\title{Half-filled Kondo lattice on the honeycomb lattice}
\author{Yin Zhong\inst{1,}\thanks{\emph{Electronic address:} zhongy05@hotmail.com} \and Ke Liu\inst{2} \and Yu-Feng Wang\inst{1} \and Yong-Qiang Wang\inst{2} \and Hong-Gang Luo \inst{1,}\inst{3,}\thanks{\emph{Electronic address:} luohg@lzu.edu.cn}%
}                     % Do not remove
\institute{Center for Interdisciplinary Studies $\&$ Key Laboratory for
Magnetism and Magnetic Materials of the MoE, Lanzhou University, Lanzhou 730000, China \and Institute of Theoretical Physics, Lanzhou University, Lanzhou 730000, China \and Beijing Computational Science Research Center, Beijing 100084, China}
\date{Received: date / Revised version: date}
\abstract{
The unique linear density of state around the Dirac points for the honeycomb lattice
brings much novel features in strongly correlated models. Here we study the ground-state
phase diagram of the Kondo lattice model on the honeycomb lattice at half-filling
by using an extended mean-field theory. By treating magnetic interaction and Kondo screening
on an equal footing, it is found that besides a trivial discontinuous first-order
quantum phase transition between well-defined Kondo insulator and antiferromagnetic insulating state,
there can exist a wide coexistence region with both Kondo screening and antiferromagnetic orders
in the intermediate coupling regime. In addition, the stability of Kondo insulator requires a minimum
strength of the Kondo coupling. These features are attributed to the linear density of state,
which are absent in the square lattice. Furthermore, fluctuation effect beyond the mean-field decoupling
is analyzed and the corresponding antiferromagnetic spin-density-wave transition falls into the $O(3)$ universal class.
Comparatively, we also discuss the Kondo necklace and the Kane-Mele-Kondo (KMK) lattice models on the same lattice.
Interestingly, it is found that the topological insulating state is unstable to the usual antiferromagnetic
ordered states at half-filling for the KMK model. The present work may be helpful for further study
on the interplay between conduction electrons and the densely localized spins on the honeycomb lattice.
 % end of PACS codes
} %end of abstract
\maketitle
\section{Introduction}\label{intro}
It is still a challenge to understand the emergent quantum phases and corresponding quantum criticality in heavy fermion systems\cite{Sachdev2011,Rosch,Vojta,Custers1,Custers2,Matsumoto,Senthil2003,Senthil2004,Pepin2005,Kim2010,Senthil2010,Zhong2012e}. To attack this challenging problem, the so-called Kondo lattice model is introduced, which is believed to capture the nature of interplay between Kondo screening and the magnetic interaction, namely, the Ruderman-Kittel-Kasuya-Yosida exchange interaction, mediated by conduction electrons among localized spins\cite{Tsunetsugu}. The former effect favors a nonmagnetic spin singlet state in strong coupling limit while the latter tends to stabilize usual magnetic ordered states in weak coupling limit. There seems to exist a quantum phase transition or even a coexistence regime between these two kinds of well-defined states\cite{Lacroix,Zhang2000,Capponi,Watanabe,Zhang2010,Zhang2011}, however, a more radical critical phase may have been observed in $YbRh_{2}(Si_{0.95}Ge_{0.05})_{2}$\cite{Custers1,Custers2010}.

The nature of these mentioned phenomena has been a long standing but controversial issue since the work of Doniach\cite{Doniach}. It is noted that an extended mean-field theory treating the magnetic interaction and the Kondo screening on an equal footing, has predicted a coexistence regime of the disorder Kondo singlet and the antiferromagnetic ordered state with small staggered magnetization and partially screened local moments in the intermediate coupling\cite{Zhang2000}. This coexistence has been confirmed by sophisticated quantum Monte Carlo (QMC) simulation without notorious `minus sign' problem\cite{Capponi,Watanabe}. Therefore, it is fair to say that such a mean-field theory could provide reliable physical results in the intermediate and strong Kondo coupling regimes when the unconventional quantum phases, e.g., quantum spin liquids, are irrelevant in the related problems\cite{Wen,Saremi}.

Recently, much attention has been focused in strongly correlated physics on the honeycomb lattice since the low energy excitations are described by relativistic Dirac fermions around distinct Dirac points rather than usual non-relativistic Landau quasiparticle near Fermi surface\cite{Neto,Kotov2010,Meng,Herbut2006,Clark,Mezzacapo2012,Zhong2012,Hasan2010,Qi2011,Rachel,Hohenadler2011,Ruegg2012}.

So far, most of the theoretical studies in this active field focus on Hubbard model or its derivative with spin-orbit coupling, the Kane-Mele-Hubbard\cite{Kotov2010,Meng,Clark,Mezzacapo2012,Rachel,Hohenadler2011,Ruegg2012}. How the conduction electrons interplay with the densely localized spins on the honeycomb lattice, which could be encoded in terms of the Kondo lattice model in principle, is largely an open problem\cite{Feng}.

\begin{figure}
\resizebox{0.5\textwidth}{!}{%
  \includegraphics{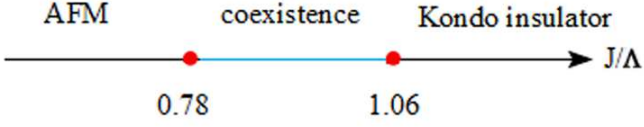}%
}
\caption{A possible ground-state phase diagram of the Kondo lattice model on the honeycomb lattice at half-filling. AFM refers the antiferromagnetic ordered insulating state in weak Kondo coupling while Kondo insulator appears in strong coupling. In the intermediate regime, a coexistence region is found with both Kondo screening and antiferromagnetic orders.}
\label{fig:1}
\end{figure}

In this work, we try to uncover the ground-state phase diagram of the Kondo lattice model on the honeycomb lattice at half-filling using the mentioned extended mean-field theory for an anisotropic Kondo lattice model. Our main results are systematically summarized in Fig.~\ref{fig:1}. For a weak Kondo coupling an antiferromagnetic ordered insulating state appears, and in strong coupling limit the Kondo insulator is the ground state, instead. In the intermediate coupling regime, we find that a wide coexistence regime with both Kondo screening and long-range antiferromagnetic order exists beside a trivial discontinuous first-order quantum phase transition. The appearance of such wide coexistence region is attributed to the unique linear density of state on the honeycomb lattice at half-filling in contrast to the case of square lattice. It is expected such a coexistence region could be realized by experiments of ultra-cold atoms on the honeycomb optical lattices and might be found by quantum Monte Carlo simulation in future.

Additionally, when fluctuation effect beyond the mean-field decoupling is introduced, the the corresponding antiferromagnetic spin-density-wave transition falls into the $O(3)$ universal class. We also find that the stability of Kondo insulator requires a minimum strength of the Kondo coupling, which is also the result of the magic linear density of state. Otherwise, the ground state of the Kondo lattice is a trivial decoupled state, which may evolves into the fractionalized Fermi liquid state with appropriate magnetic frustration interaction\cite{Senthil2003,Senthil2004,Saremi}.
In addition, the ground-state phase diagram of the Kondo necklace \cite{Doniach} and the Kane-Mele-Kondo lattice models \cite{Feng} are also discussed. Particularly, for the latter model, we find the topological insulating state found by Feng \textit{et.al.} is unstable to the usual antiferromagnetic ordered states at half-filling. We hope the present work may be helpful for further studies on the interplay between conduction electrons and the densely localized spins for the honeycomb lattice.

The remainder of the present paper is organized as follows. In Sec. \ref{sec2}, the Kondo lattice model on the honeycomb lattice is introduced and the mean-field decoupling is performed. In Sec. \ref{sec3}, the mean-field solution for the ground-state is studied and the ground-state phase diagram is established. Furthermore, fluctuation effect beyond the mean-field decoupling is analyzed in Sec. \ref{sec4}. As byproducts, the Kondo necklace model and the Kane-Mele-Kondo lattice model are also discussed in Sec. \ref{sec5}. Finally, a concise conclusion is devoted to Sec. \ref{sec6}.

\section{The Kondo lattice model on the honeycomb lattice and mean-field decoupling} \label{sec2}
The model we considered is the anisotropic Kondo lattice model defined on the honeycomb lattice at half-filling\cite{Zhang2000},
\begin{eqnarray}
&&H=H_{t}+H_{\parallel}+H_{\perp},\nonumber\\
&&H_{t}=-t\sum_{\langle ij\rangle \sigma}(c_{i\sigma}^{\dag}c_{j\sigma}+c_{j\sigma}^{\dag}c_{i\sigma}),\nonumber\\
&&H_{\parallel}=\frac{J_{\parallel}}{4}\sum_{i}(c_{i\uparrow}^{\dag}c_{i\uparrow}-c_{i\downarrow}^{\dag}c_{i\downarrow})(d_{i\uparrow}^{\dag}d_{i\uparrow}-d_{i\downarrow}^{\dag}d_{i\downarrow}),\nonumber\\
&&H_{\perp}=\frac{J_{\perp}}{2}\sum_{i}(c_{i\uparrow}^{\dag}c_{i\downarrow}d_{i\downarrow}^{\dag}d_{i\uparrow}+c_{i\downarrow}^{\dag}c_{i\uparrow}d_{i\uparrow}^{\dag}d_{i\downarrow}), \label{eq1}
\end{eqnarray}
where $H_{t}$ describes conduction electrons hopping between nearest-neighbor sites and the pseudofermion representation for local spins has been utilized as $S_{i}^{\alpha}=\frac{1}{2}\sum_{\sigma\sigma'}d_{i\sigma}^{\dag}\tau_{\sigma\sigma'}^{\alpha}d_{i\sigma'}$ with $\tau^{\alpha}$ being usual Pauli matrix and a local constraint $d_{i\uparrow}^{\dag}d_{i\uparrow}+d_{i\downarrow}^{\dag}d_{i\downarrow}=1$ enforced in every site. $H_{\parallel}$ denotes the magnetic instability due to the polarization of conduction electrons by local spins while $H_{\perp}$ describes the local Kondo screening effect resulting from spin-flip scattering process of conduction electrons by local moments. The Kondo screening effect has also been formulated in the $1/N$ expansion but magnetic instability is difficult to treat in such a framework\cite{Doniach1987}.

Now, using the mean-field decoupling introduced by Zhang and Yu \cite{Zhang2000} for the longitudinal and transverse interaction term $H_{\parallel}$, $H_{\perp}$, respectively, it is straightforward to obtain a mean-field Hamiltonian
\begin{eqnarray}
&&H_{MF}=H_{t}+H_{\parallel}^{MF}+H_{\perp}^{MF}+E_{0},\nonumber\\
&&H_{t}=-t\sum_{k\sigma}(f(k)c_{kA\sigma}^{\dag}c_{kB\sigma}+f^{\star}(k)c_{kB\sigma}^{\dag}c_{kA\sigma}),\nonumber\\
&&H_{\parallel}^{MF}=\frac{J_{\parallel}}{2}\sum_{k\sigma}[\sigma(-m_{c}d_{kA\sigma}^{\dag}d_{kA\sigma}+m_{d}c_{kA\sigma}^{\dag}c_{kA\sigma})\nonumber\\
&& \hspace{2cm} - (A\rightarrow B)]\nonumber\\
&&H_{\perp}^{MF}=\frac{J_{\perp}V}{2}\sum_{k\sigma}(c_{kA\sigma}^{\dag}d_{kA\sigma}+c_{kB\sigma}^{\dag}d_{kB\sigma}+h.c.),\nonumber\\
&&E_{0}=N_{s}(2J_{\parallel}m_{d}m_{c}+J_{\perp}V^{2}), \label{eq2}
\end{eqnarray}
where we have defined several mean-field parameters as $\langle d_{iA\uparrow}^{\dag}d_{iA\uparrow}-d_{iA\downarrow}^{\dag}d_{iA\downarrow}\rangle = 2m_{d}$,
$\langle d_{iB\uparrow}^{\dag}d_{iB\uparrow}-d_{iB\downarrow}^{\dag}d_{iB\downarrow}\rangle = -2m_{d}$, $\langle c_{iA\uparrow}^{\dag}c_{iA\uparrow}-c_{iA\downarrow}^{\dag}c_{iA\downarrow}\rangle = -2m_{c}$,
$\langle c_{iB\uparrow}^{\dag}c_{iB\uparrow}-c_{iB\downarrow}^{\dag}c_{iB\downarrow}\rangle = 2m_{c}$ and
$-V=\langle c_{i\uparrow}^{\dag}d_{i\uparrow}+d_{i\downarrow}^{\dag}c_{i\downarrow}\rangle =
\langle c_{i\downarrow}^{\dag}d_{i\downarrow}+d_{i\uparrow}^{\dag}c_{i\uparrow}\rangle $ with $f(k)=e^{-ik_{x}}+2e^{ik_{x}/2}\cos\sqrt{3}k_{y}/2$ and $A$, $B$ representing two nonequivalent sublattices. It can be seen that $m_{d},m_{c}$ corresponds to magnetization of local spins and conduction electrons, respectively, while non-vanishing $V$ denotes the onset of Kondo screening effect. For simplicity, we have assumed that the magnetic instability leads to the collinear
antiferromagnetic ordered state where the order parameter (staggered magnetization) has opposite spin orientation in the two sublattice.
Besides, since we are considering a half-filled lattice, the local constraint has been safely neglected at the present mean-field level with chemical potential setting to zero\cite{Zhang2000}.

Diagonalizing the above mean-field Hamiltonian, the four quasiparticle bands are obtained as
\begin{eqnarray}
E_{\pm\pm}(k)=\pm\frac{1}{\sqrt{2}}\sqrt{F_{1}(k)\pm F_{2}(k)}\label{eq3}
\end{eqnarray}
with $F_{1}(k)=J_{\parallel}^{2}(m_{d}^{2}+m_{c}^{2})/4+J_{\perp}^{2}V^{2}/2+t^{2}|f(k)|^{2}$
and $F_{2}(k)=[F_{1}(k)^{2}-4(J_{\parallel}^{4}m_{d}^{2}m_{c}^{2}/16+J_{\parallel}^{2}m_{d}m_{c}J_{\perp}^{2}V^{2}/8+J_{\perp}^{4}V^{4}/16)-J_{\parallel}^{2}m_{c}^{2}t^{2}|f(k)|^{2}]^{\frac{1}{2}}$.
Therefore, the wanted ground-state energy at half-filling is given by
\begin{eqnarray}
E_{g}=\sum_{k}(E_{--}(k)+E_{-+}(k))+E_{0}.\label{eq4}
\end{eqnarray}

\section{The mean-field solution for ground-state}\label{sec3}
Firstly, we proceed to discuss two simple but physically interesting limits for Kondo coupling $J_{\parallel}$ and $J_{\perp}$, which correspond to the antiferromagnetic ordered state ($J_{\parallel}\gg J_{\perp}$) and Kondo insulating state ($J_{\parallel}\ll J_{\perp}$), respectively\cite{Tsunetsugu}.

\subsection{The antiferromagnetic insulating state}
For the case with $J_{\parallel}\gg J_{\perp}$, in general, one expects the antiferromagnetic ordered state to be the stable ground-state of Kondo lattice model on the honeycomb lattice due to its bipartite feature\cite{Tsunetsugu}. To study the possible antiferromagnetic ordered state, making use of the ground-state energy (Eq.\ref{eq4}) with assuming no Kondo screening existing ($V=0$) and with the help of the quasiparticle spectrum (Eq.\ref{eq3}), we can easily derive ground-state energy of the antiferromagnetic ordered state per site as
\begin{eqnarray}
E_{g}^{AFM}=J_{\parallel}m_{c}(2m_{d}-1)-\frac{1}{N_{s}}\sum_{k}\sqrt{J_{\parallel}^{2}m_{d}^{2}+4t^{2}|f(k)|^{2}}.\nonumber
\end{eqnarray}
and two self-consistent equations from minimizing $E_{g}^{AFM}$ with respect to magnetization $m_{d}$ and $m_{c}$, respectively.
\begin{eqnarray}
&&J_{\parallel}m_{c}(2m_{d}-1)=0,\nonumber\\
&&2J_{\parallel}m_{c}-\frac{J_{\parallel}^{2}m_{d}}{2\Lambda^{2}}[\sqrt{4\Lambda^{2}+J_{\parallel}^{2}m_{d}^{2}}-J_{\parallel}m_{d}]=0\nonumber
\end{eqnarray}
where we have used a simplified linear density of state (DOS) $\rho(\varepsilon)=|\varepsilon|/\Lambda^{2}$ when transforming the summation over momentum $k$ into integral on energy $\varepsilon$ with $\Lambda$ being high-energy cutoff\cite{Neto}. Thus, $t|f(k)|$ can be replaced by $|\varepsilon|$ to simplify corresponding calculations. (One may wonder if the true honeycomb dispersion is used, to what extent the following results change. Since $\Lambda=t\sqrt{3\pi/\sqrt{3}}\simeq2.33t$, for example, the first-order phase transition point will shift from $2.05t$($0.88\Lambda$) to $2.16t$ and only quantitative changes will be found when one uses the true honeycomb dispersion. Other physical quantities will also be modified but no significant changes appear.)

From these two equations, one obtains $m_{d}=1/2$ and $m_{c}=\frac{J_{\parallel}}{8\Lambda^{2}}[\sqrt{4\Lambda^{2}+J_{\parallel}^{2}/4}-J_{\parallel}/2]$. Meanwhile, the low-lying quasiparticle excitations in the antiferromagnetic ordered state has the energy $E_{\pm+}(k)=\pm\sqrt{t^{2}|f(k)|^{2}+J_{\parallel}^{2}/16}$ and $E_{\pm-}(k)=\pm J_{\parallel}m_{c}/2$ with an apparent gap $J_{\parallel}/4$ around the Dirac points where $f(k)=0$. Obviously, such gap for the quasiparticle results from interplay between conduction electrons and the antiferromagnetic background formed by unscreened local spins. Thus, we conclude that the antiferromagnetic ordered state we obtained is indeed an insulating state with fully polarized local spins ($m_{d}=1/2$) while conduction electrons only partially polarize ($m_{c}<1/2$). This feature is similar to the previous study on square lattice, thus confirms the validity of our current treatment\cite{Zhang2000}.
\begin{figure}
\resizebox{0.5\textwidth}{!}{%
  \includegraphics{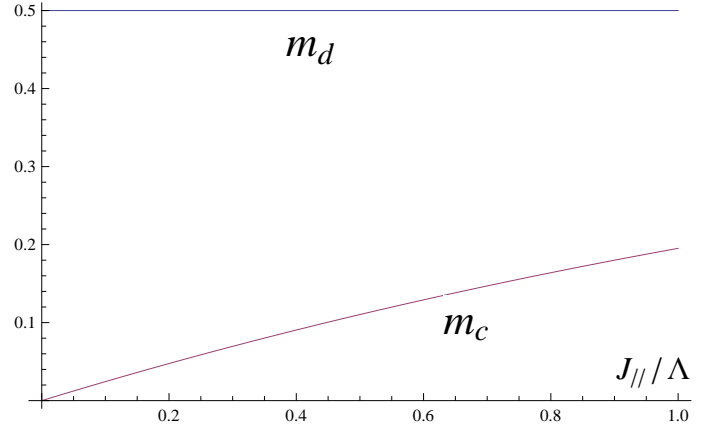}%
}
\caption{Order parameters $m_{d}$ and $m_{c}$ in the antiferromagnetic ordered state.}
\label{fig:2}
\end{figure}

We also note that the local spins are localized in their own site (They have no dispersion at all.) and no heavy electrons are formed in the present antiferromagnetic ordered state.

\subsection{The Kondo insulating state}
Another interesting case appears when $J_{\parallel}\ll J_{\perp}$. It is natural to expect that a Kondo insulating  state arises in this situation for half-filling\cite{Tsunetsugu,Zhang2000}. Following the same methology of treating antiferromagnetic insulating state, we can get the ground-state energy per site for the expected Kondo insulating state with $V\neq0$ but no magnetic orders $m_{d}=m_{c}=0$
\begin{eqnarray}
E_{g}^{Kondo}=J_{\perp}V^{2}-\frac{4}{3\Lambda^{2}}[(\Lambda^{2}+J_{\perp}^{2}V^{2})^{3/2}-J_{\perp}^{3}V^{3}]\nonumber
\end{eqnarray}
and the quasiparticle excitations energy
\begin{eqnarray}
E_{\pm\pm}(k)=\pm\frac{1}{2}[\sqrt{(t|f(k)|)^{2}+J_{\perp}^{2}V^{2}}\pm t|f(k)|]\nonumber
\end{eqnarray}

Minimizing $E_{g}^{Kondo}$ with respect to Kondo hybridization parameter $V$, we obtain $V=1-\Lambda^{2}/(4J_{\perp}^{2})$ which implies a critical coupling $J_{\perp}^{c}=1/2$ corresponding to vanishing $V$. It is noted that $V\propto(J_{\perp}-J_{\perp}^{c})$ (with critical exponent $\beta=1$) in contrast to usual mean-field result $\beta=1/2$, and this can be attributed to the low-energy linear DOS of conduction electrons on the honeycomb lattice at half-filling.
Similar critical behavior for onset of Kondo screening on the honeycomb lattice has been obtained in the study of Kondo breakdown mechanism as well\cite{Saremi}. As a matter of fact, the existence of the critical coupling $J_{\perp}^{c}=1/2$ results from the competition between the Kondo insulating state and the trivial decoupled state where $V=m_{d}=m_{c}=0$, its ground-state energy $E_{g}^{0}=-4\Lambda/3$ comes solely from free conduction electrons. Comparing $E_{g}^{0}$ and $E_{g}^{Kondo}$, one clearly recovers the critical coupling $J_{\perp}^{c}$, which justifies the above simple picture.
\begin{figure}
\resizebox{0.5\textwidth}{!}{%
  \includegraphics{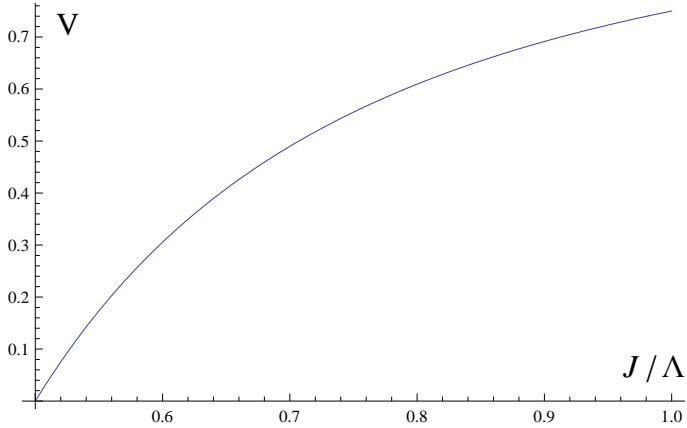}%
}
\caption{Kondo hybridization parameter $V$ in the Kondo insulating state.}
\label{fig:3}
\end{figure}
One may also interest the physical properties of quasiparticles in the Kondo insulating state. Apparently, at half-filling the quasiparticles are gapped by the hybridizing between conduction electrons and local spins via the Kondo screening ($V\neq0$) and such quasiparticles are heavy fermions since the the specific heat coefficient ($\gamma^{\star}$) is larger than the case with only free conduction electrons. ($\gamma^{\star}/\gamma\simeq1/V^{2}$ with $V<1$)

\subsection{The nature of the trivial decoupled state}
We should emphasize that at the current mean-field level, the mentioned trivial decoupled state has $V=m_{d}=m_{c}=0$, which just indicates that conduction electrons are decoupled from local spins and behave as a usual Dirac semimetal while those local spins do not develop any magnetic orders. In fact, when the direct Heisenberg exchange interaction between those local spins is introduced, we expect the
trivial decoupled state should evolves into the `fractionalized Fermi liquid state' as studied in Refs.
\cite{Senthil2003,Senthil2004,Saremi}, where local spins form a highly-entangled quantum spin liquid and conduction electrons form the Dirac semimetal. However, because the mean-field decoupling used in the present work is unable to capture the essential feature of quantum spin liquids, thus we do not discuss this issue in the present framework.

Moreover, it is noted that the Kondo insulating state is unstable to the decoupled state when $J_{\perp}<J_{\perp}^{c}=1/2$. One may wonder whether a trivial decoupled state appears between the Kondo insulating state and the antiferromagnetic insulating state. It is easy to see the ground-state energy of the antiferromagnetic state $E_{g}^{AFM}$ is always lower than $E_{g}^{0}$ for any positive coupling $J_{\parallel}$. Therefore, a trivial disordered state in intermediate coupling seems unfavorable based on our current mean-field treatment. It seems that the instability of the decoupled disordered state to antiferromagnetic ordered state is a general feature for the Kondo lattice model at half-filling.

\subsection{Possible first-order quantum phase transition}
Since the possible trivial decoupled state can be precluded due to the discussion in last subsection, we may consider a possible quantum critical point (QCP) between Kondo insulating state and the antiferromagnetic state and its position can be determined by comparing the ground-state energies of $E_{g}^{AFM}$ and $E_{g}^{Kondo}$. For physically interesting case with $J_{\perp}=J_{\parallel}=J$, we have
\begin{equation}
\frac{x^2}{6}\left[\left(\frac{4}{x^2} + \frac{1}{4}\right)^\frac{3}{2} - \frac{1}{8}\right] = -y + \frac{4x^2}{3}\left[\left(\frac{1}{x^2} + y^2\right)^{\frac{3}{2}} - y^3\right],\label{eq5}
\end{equation}
where $x = J/\Lambda$ and $y = 1-\frac{1}{4x^2}$.
The position of the QCP is readily obtained as $J_{c}=0.88\Lambda$, numerically. However, in fact, one can check that this putative QCP is in fact a first-order quantum phase transition point when comparing the first-order derivative
of $E_{g}^{Kondo}$ and $E_{g}^{AFM}$ with respect to the Kondo coupling $J$. Thus, we do not expect radical critical behaviors near such first-order quantum phase transition point in the spirit of Landau-Ginzburg-Wilson paradigm\cite{Sachdev2011}. Additionally, it is noted that such first-order quantum phase transition has also been obtained on the square lattice \cite{note} and we suspect this feature may be generic for conventional mean-field treatment of Kondo lattice models according to standard Landau-Ginzburg phase transition theory\cite{Zhang2000}. However, we point out that it is a subtle issue to compare the results of a first-order quantum transition with numerical simulations, particularly when the first-order transition is a weak one\cite{Continentino}.
\begin{figure}
\resizebox{0.5\textwidth}{!}{%
  \includegraphics{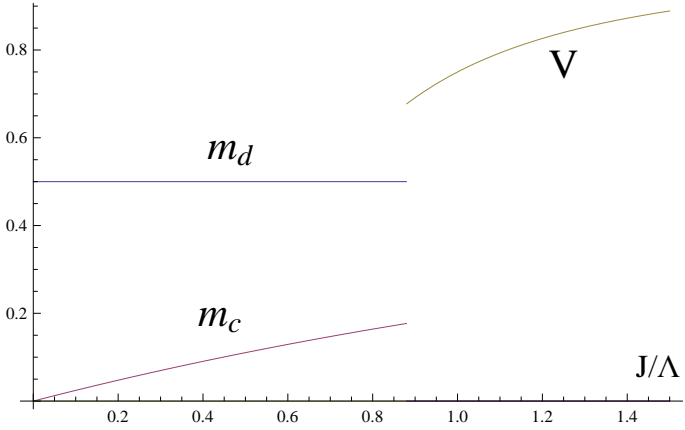}%
}
\caption{Order parameters $m_{d}$, $m_{c}$ and $V$ around the first-order quantum phase transition point $J/\Lambda=0.88$.}
\label{fig:4}
\end{figure}
\subsection{Possible coexistence region of the Kondo insulating state and the antiferromagnetic ordered state}
Generically, in the intermediate coupling regime, a possible coexistence region of the Kondo insulating state and the antiferromagnetic ordered state cannot be excluded\cite{Zhang2000}. For searching such possibility, we have to resort to the full formulism of ground-state energy $E_{g}$ (Eq.\ref{eq4}) and three self-consistent equations derived from $\frac{\partial E_{g}}{\partial V}=\frac{\partial E_{g}}{\partial m_{d}}=\frac{\partial E_{g}}{\partial m_{c}}=0$.
Since the possible coexistence region has two boundaries, which correspond to the onset of Kondo screening ($V\neq0$) in antiferromagnetic insulating state and antiferromagnetic order ($m_{c},m_{d}\neq0$) arising in Kondo insulator, we can derive two equations for these two distinct boundaries.

The first boundary which means the antiferromagnetic order ($m_{c},m_{d}\neq0$) arising in Kondo insulator can be obtained by solving the following equation
\begin{equation}
y^{2}=\frac{2}{3}\left(1 - x^2 - \frac{1}{16x^{2}}\right)\left(\frac{1}{x^2} + y^2\right)^\frac{1}{2} +\frac{2}{3}x^2 y^{3},\label{eq6}
\end{equation}
where we have utilized the expression for the Kondo hybridization parameter $V=1-\Lambda^{2}/(4J_{\perp}^{2})$ and assuming the isotropic condition ($J_{\perp}=J_{\parallel}=J$). Solving Eq.\ref{eq6} numerically, one finds the first boundary located in $J_{c1}=0.78\Lambda$ which is smaller than first-order quantum phase transition point ($J_{c}=0.88\Lambda$), thus the expected coexistent scenario is stable to the previous discontinuous first-order quantum phase transition.

For the second boundary, which corresponds to the onset of Kondo screening ($V\neq0$) in antiferromagnetic state, one
can also derive similar equation like Eq.(\ref{eq6}) to determine the location of the boundary.
\begin{eqnarray}
\frac{2\Lambda^{2}}{J}=\int_{0}^{\Lambda}d\varepsilon \varepsilon[P(\varepsilon)+Q(\varepsilon)+(P(\varepsilon)-Q(\varepsilon))W(\varepsilon)],\label{eq7}
\end{eqnarray}
where we have defined three auxiliary functions $P(\varepsilon)=1/\sqrt{\varepsilon^2+J^{2}/16}$, $Q(\varepsilon)=2/Jm_{c}$ and $W(\varepsilon)=[\varepsilon^2+J^{2}/16+J^{2}(m_{c}^{2}-m_{c})/4]/[\varepsilon^2+J^{2}(1/4-m_{c}^{2})/4]$ with $m_{c}=J_{\parallel}[(4\Lambda^{2}+J_{\parallel}^{2}/4)-J_{\parallel}/2]/8$. Then, one obtains the location of the second boundary $J_{c2}=1.06\Lambda$ by numerically solving Eq.(\ref{eq7}). It is noted that $J_{c2}=1.06\Lambda>J_{c}=0.88\Lambda$, which implies the system indeed tends to form a coexistent state with Kondo screening and antiferromagnetic orders.

Therefore, we may conclude that there may exist a coexistence regime ($J_{c1}=0.78\Lambda<J<J_{c2}=1.06\Lambda$) in the intermediate coupling regime based on the discussion in two phase boundaries for the isotropic case ($J_{\perp}=J_{\parallel}=J$). When comparing to the square lattice, we find the range of the coexistence regime on the honeycomb lattice is obviously larger than the one on the square lattice with $J_{c1}=0.56\Lambda<J<J_{c2}=0.62\Lambda$\cite{Zhang2000}. We suspect this difference may result from unique linear DOS on the honeycomb lattice at half-filling, which lowers the energy penalty of coexistence. Besides, in such a coexistence regime the quasiparticle excitation is completely gapped due to the antiferromagnetic order while the staggered magnetization of both local spins and conduction electrons should be smaller than the one in the antiferromagnetic insulating state.

\section{Fluctuation effect beyond the mean-field decoupling}\label{sec4}
In previous section, we have studied the mean-field solution for the ground-states of the half-filled Kondo lattice model on the honeycomb lattice, here, we give some arguments on the fluctuation effect beyond the mean-field decoupling.

Since the most interesting result of mean-field decoupling in Sec. \ref{sec3} is the phase diagram Fig.~\ref{fig:1},
we will mainly focus on states in this phase diagram.

First, for the Kondo insulator state, fluctuation correction will come from the local constraint term ($d_{i\uparrow}^{\dag}d_{i\uparrow}+d_{i\downarrow}^{\dag}d_{i\downarrow}=1$), which has been neglected at the previous mean-field decoupling, and non-vanishing Kondo screening 'order parameter' field $V$ ($V$ should be treated as a dynamical field if one wants to include its effect into the fluctuation correction.) as what is well-known in the studies of $1/N$ expansion\cite{Doniach1987}. Apparently, these two terms do not contribute singular corrections but only slightly renormalize the energy band of the quasiparticles and introducing weak interactions between those quasiparticles, thus the Kondo insulator state is stable in this case.

When one considers fluctuation correction in the antiferromagnetic insulating state, one may have an effective $\varphi^{4}$ theory in his mind, which describes the pure fluctuating antiferromagnetic order
\begin{eqnarray}
S_{\varphi}=\int d^{2}xd\tau\{(\partial_{\tau}\vec{\varphi})^{2}+(\nabla\vec{\varphi})^{2}+r\vec{\varphi}^{2}+u(\vec{\varphi}^{2})^{2}\}\label{eq8}
\end{eqnarray}
where $r$, $u$ are both phenomenological parameters and $\vec{\varphi}$ denotes the fluctuating antiferromagnetic order. Obviously, when fluctuation is introduced, local spin cannot fully polarize ($m_{d}<1/2$) and the only active actor in low-energy limit is the Goldstone modes (here the antiferromagnetic spin-density-wave) from the spontaneously breaking spin-rotation symmetry in antiferromagnetic ordered state. (Recall that the electronic quasiparticle is fully gapped in the antiferromagnetic ordered state.)

Then, in the coexistence region, fluctuations from the local constraint, Kondo screening 'order parameter' field and
the fluctuating antiferromagnetic order coexist but no qualitative behaviors are changed in compared to the case without any fluctuation correction. However, near the boundary of the coexistence region, phase transitions may occur.
For example, near the first boundary, which corresponds to the onset of antiferromagnetic order in the Kondo insulating state, there exists an antiferromagnetic spin-density-wave transition which can be described by Eq. \ref{eq8} since electronic quasiparticles are still gapped by Kondo screening in this region. Therefore, the critical behaviors of the  antiferromagnetic spin-density-wave transition should fall into the usual $3D$ $O(3)$ universal class. In contrast, for the second boundary, where Kondo screening ($V$) vanishes while antiferromagnetic order persists, we do not expect radical critical behaviors because vanishing of Kondo screening may be considered as a crossover rather than a true phase transition and electronic quasiparticles are gapped by the antiferromagnetic order.

\section{Extensions and discussions} \label{sec5}
\subsection{Kondo necklace model on honeycomb model}
We noted that, it is also interesting to study a more simplified model only including degrees of freedom for spins, namely, the Kondo necklace model on the honeycomb lattice.\cite{Doniach}
\begin{eqnarray}
H_{KN}=-t\sum_{\langle ij\rangle}(\tau_{i}^{x}\tau_{j}^{x}+\tau_{i}^{y}\tau_{j}^{y})+J\sum_{i}\vec{\tau_{i}}\cdot\vec{S_{i}}\nonumber
\end{eqnarray}
where  $\tau_{i}^{x},\tau_{i}^{y}$ represents the spin degrees of freedom coming from original conduction electrons and $\vec{S}_{i}$ denoting local spins. However, we should remind the reader that the Kondo necklace model cannot be derived from the original Kondo lattice model at half-filling but could only be considered as a phenomenological model devised for studying the low-lying spin excitations.

\begin{figure}
\resizebox{0.5\textwidth}{!}{%
  \includegraphics{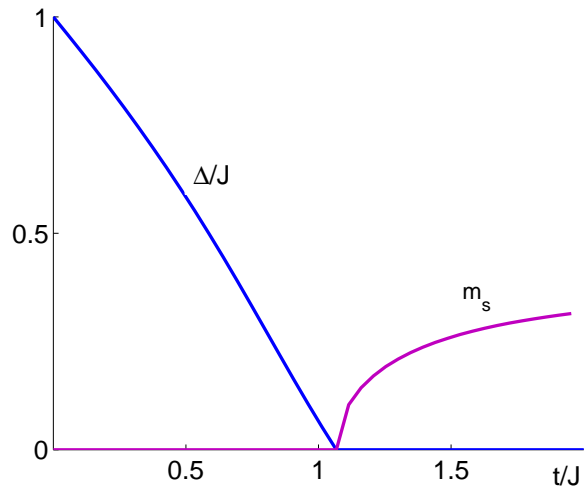}%
}
\caption{Mean-field results of the quasiparticle gap $\Delta$ and magnetization $m_{s}$ for the Kondo necklace model. Note that the order parameter magnetization $m_{s}$ vanishes continuously when approaching the critical point  $t/J=1.0673$, thus a second-order quantum phase transition is obtained and first-order quantum phase transition or coexistent region is excluded.}
\label{fig:5}
\end{figure}

Unfortunately, the present mean-field decoupling approach in previous section is not cheap for studying such a spin model but the so-called bond-operator representation could be a useful tool in this case\cite{Gu}. When the bond-operator mean-field approximation is used for the Kondo necklace model on the honeycomb lattice, we find a continuous second-order quantum phase transition between the Kondo insulating phase and antiferromagnetic insulating state, which resembles the results on the square lattice\cite{Gu}. Mean-field results of the quasiparticle gap $\Delta$ and magnetization $m_{s}$ for the Kondo necklace model is shown in Fig.~\ref{fig:5}. The existence of quasiparticle gap denotes that there is a Kondo insulating phase if $t/J<1.0673$. The antiferromagnetic insulating state is clearly seen with a non-zero magnetization $m_{s}$ when $t/J>1.0673$.

The above result is not surprising since the bond-operator representation introduces some auxiliary bosons to represent the spin operators and the low-energy behavior of those auxiliary bosons cannot be influenced by the Dirac points. Thus, the most important feature of honeycomb lattice at half-filling, the structure of Dirac points (cones), is lost in the bond-operator representation and it seems not reliable to compare the results of the Kondo necklace model to the original Kondo lattice model.

\subsection{Kane-Mele-Kondo lattice model}
Moreover, if a spin-orbit coupling is introduced in our present model as the case in Feng et.al. \cite{Feng}, one could study the subtle interplay among the antiferromagnetic order, Kondo screening and the symmetry-protected topological states (such as the $2D$ topological insulator). The Kane-Mele-Kondo lattice model used by Feng et.al. reads as follows
\begin{eqnarray}
&&H=H_{KM}+H_{\parallel}+H_{\perp},\nonumber\\
&&H_{KM}=-t\sum_{\langle ij\rangle \sigma}(c_{i\sigma}^{\dag}c_{j\sigma}+c_{j\sigma}^{\dag}c_{i\sigma})-t'\sum_{\langle\langle ij\rangle\rangle \sigma}e^{i\varphi_{ij}}c_{i\sigma}^{\dag}\tau_{\sigma\sigma'}^{z}c_{j\sigma'},\nonumber\\
&&H_{\parallel}=\frac{J_{\parallel}}{4}\sum_{i}(c_{i\uparrow}^{\dag}c_{i\uparrow}-c_{i\downarrow}^{\dag}c_{i\downarrow})(d_{i\uparrow}^{\dag}d_{i\uparrow}-d_{i\downarrow}^{\dag}d_{i\downarrow}),\nonumber\\
&&H_{\perp}=\frac{J_{\perp}}{2}\sum_{i}(c_{i\uparrow}^{\dag}c_{i\downarrow}d_{i\downarrow}^{\dag}d_{i\uparrow}+c_{i\downarrow}^{\dag}c_{i\uparrow}d_{i\uparrow}^{\dag}d_{i\downarrow}),\nonumber
\end{eqnarray}
where $H_{KM}$ is the standard Kane-Mele Hamiltonian defined on the honeycomb lattice\cite{Kane2005}.

\begin{figure}
\resizebox{0.5\textwidth}{!}{%
  \includegraphics{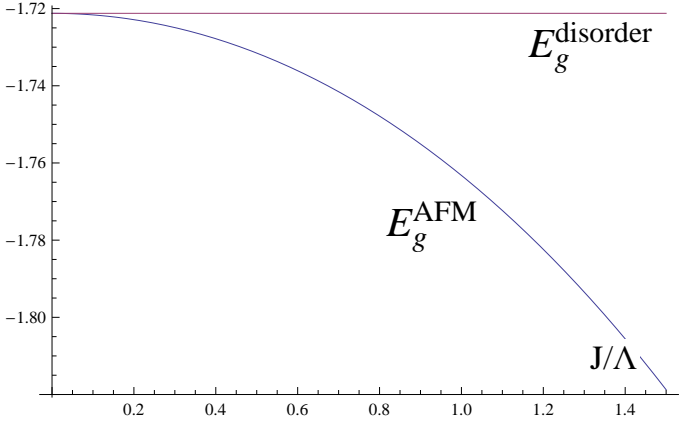}%
}
\caption{Ground-state energy per site for the topological insulating state $E^{disorder}_{g}$ and the usual antiferromagnetic state $E^{AFM}_{g}$ with $t'=0.1$.}
\label{fig:6}
\end{figure}

\begin{figure}
\resizebox{0.5\textwidth}{!}{%
  \includegraphics{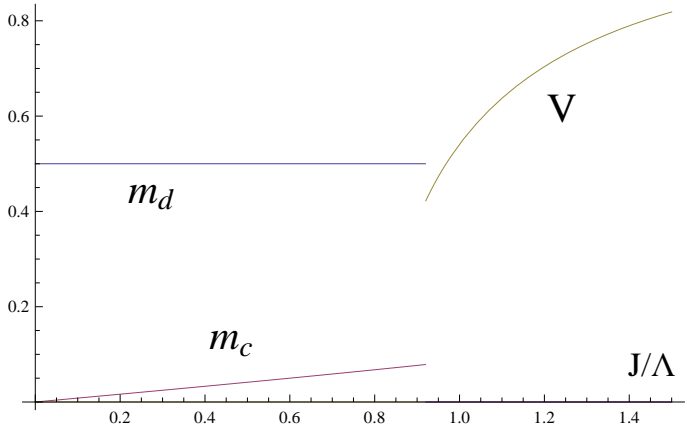}%
}
\caption{Order parameters $m_{d}$, $m_{c}$ and $V$ in the mean-field solution for the Kane-Mele-Kondo lattice model with $t'=0.1$.}
\label{fig:7}
\end{figure}

In terms of the mean-field decoupling described in previous sections, we find that at half-filling the topological insulating state ($V=m_{c}=m_{d}=0$) always has higher ground-state energy than the usual antiferromagnetic state ($V=0,m_{c}=m_{d}\neq0$) in spite of the fact that the topological insulating state is more stable than the trivial Kondo insulating state in weak Kondo coupling regime. In Fig.~\ref{fig:6}, we show the ground-state energy per site for the topological insulating state $E^{disorder}_{g}$ and the usual antiferromagnetic state $E^{AFM}_{g}$ and obviously the antiferromagnetic state always has lower energy than the topological insulating state. Order parameters $m_{d}$, $m_{c}$ and $V$ are also shown in Fig.~\ref{fig:7} with $t'=0.1$. Therefore, based on our mean-field results, at least at half-filling, the quantum phase transition studied in Ref.\cite{Feng} might give way to a trivial first-order transition between the antiferromagnetic state and Kondo insulating state if no other kinds of interaction or hopping terms are added. Further sophisticated numerical studies are eagerly desired to clarify the true ground-state of the mentioned Kane-Mele-Kondo lattice model.

\subsection{Relation to impurity Kondo model with a pseudo-gap DOS and calculations from the dynamical mean-field theory}
The impurity Kondo model with a pseudo-gap DOS has been well studied in literatures \cite{Withoff,Ingersent,Uchoa2011}. The main difference from the usual Kondo model is the existence of the critical value for the Kondo coupling $J$ in the model with a pseudo-gap DOS. (Recall that no critical value of Kondo coupling is required in the usual Kondo model to form the Kondo singlet state.) This means that only if $J>J_{critical}$ ($J_{critical}$ being the critical Kondo coupling), the ground-state will be a Kondo singlet state. As for the half-filled Kondo lattice model on the honeycomb lattice, since the density of state of the quasiparticles also has the pseudo-gap behavior, there is a critical Kondo coupling to form the Kondo insulting state as studied in Sec. \ref{sec3}. It is interesting to note that in contrast to usual mean-field result $\beta=1/2$, the low-energy linear DOS (pseudo-gap DOS) of conduction electrons on the honeycomb lattice at half-filling leads to $V\propto(J_{\perp}-J_{\perp}^{c})$ (with critical exponent $\beta=1$). In addition, We note that the comparison of the impurity Kondo model with a pseudo-gap DOS to the Kondo lattice model on the honeycomb lattice with spin-orbit coupling is detailed discussed by Feng et.al. \cite{Feng} and we refer reader to their paper.

There are various calculations on the usual Kondo lattice model by using the dynamical mean-field theory (DMFT)\cite{Otsuki,Hoshino,Bercx}. However, we find that those calculations are mainly focused on the square lattice and to our knowledge, the Kondo lattice model in the honeycomb lattice has not been studied by DMFT till now. Therefore, we cannot compare our results to the calculations from DMFT.

\section{Conclusion}\label{sec6}
In summary, we have obtained the ground-state phase diagram of the Kondo lattice model on the honeycomb lattice at half-filling using extended mean-field decoupling. By treating magnetic interaction and Kondo screening on an equal footing, it is found that besides a trivial discontinuous first-order quantum phase transition between well-defined Kondo insulator and antiferromagnetic insulating state, there can exist a wide coexistence regime with both Kondo screening and antiferromagnetic orders in the intermediate coupling regime. It is expected such a coexistence regime could be realized by experiments of ultra-cold atoms on the honeycomb optical lattices and might be found by quantum Monte Carlo simulation in future\cite{Capponi,Watanabe,Bloch,Goldman,Bermudez}.

We also find that the stability of Kondo insulator to the trivial decoupled state requires a minimum strength of the Kondo coupling, which results from the magic linear density of state of half-filled honeycomb lattice near Dirac points. Furthermore, for comparison, the Kondo necklace model and Kane-Mele-Kondo lattice model are discussed as well. We hope the present work may be helpful for further studies on the interplay between conduction electrons and the densely localized spins for the honeycomb lattice.

\section{Acknowledgments}
The work was supported partly by NSFC, the Program for NCET, the Fundamental Research Funds for the Central Universities and the national program for basic research of China.

\end{document}